\documentclass{entcs} \usepackage{entcsmacro}
\usepackage{graphicx}

\usepackage{amsmath}
\usepackage{amssymb}
\usepackage{stmaryrd}

\usepackage{xspace}

\usepackage{boxedminipage}
\usepackage{rotating}
\usepackage{subfigure}

\newcommand{\union}{\cup}

\newcommand{\rond}{\circ}

\newcommand{\fleche}{\rightarrow}
\newcommand{\dessusdessous}[2]{\genfrac{}{}{0pt}{0}{#1}{#2}}

\newcommand{\MGS}{MGS\xspace}

\newcommand{\codeline}[1]{\texttt{#1}}
\newcommand{\cod}{\codeline}

\newcommand{\FVt}{\mathcal L _t}
\newcommand{\FVr}{\mathcal L _r}

\renewcommand{\annum}{2003}

\def\lastname{Cohen}
\begin{document}
\begin{frontmatter}
  \title{Typing rule-based transformations over topological collections}
 \author{Julien Cohen\thanksref{remercie}
}
  \address{LaMI, U.M.R. 8042\\ CNRS -- Universit\'e d'\'Evry Val d'Essonne\\
	523 place des Terrasses de l'Agora\\
    91000 \'Evry, France}  
\thanks[remercie]{The author is grateful to Olivier Michel and Jean-Louis Giavitto of the MGS Project for their valuable support.
}
\begin{abstract}

Pattern-matching programming is an example of a rule-based programming
style developed in  functional languages. This programming style is
intensively used in dialects of ML but is restricted to algebraic
data-types.

This restriction limits the field of application. However, as shown
by~\cite{rule02} at RULE'02, case-based function definitions can be
extended to more general data structures called \emph{topological
collections}. We show in this paper that this extension retains the
benefits of the typed discipline of the functional languages. More
precisely, we show that topological collections and the rule-based
definition of functions associated with them fit in a polytypic
extension of mini-ML where type inference is still possible.

\end{abstract}
\end{frontmatter}

\newcommand{\scouleur}{\sqsubseteq}
\newcommand{\infere}{\vdash}
\newcommand{\reduit}{\hookrightarrow}

\newcommand{\code}{\codeline}

\newcommand{\col}[2]{[#2]#1}

\newcommand{\self}{\code{self}}
\newcommand{\return}{\code{return}~}
\newcommand{\seq}{\code{seq}}
\renewcommand{\int}{\code{int}}
\newcommand{\set}{\code{set}}
\newcommand{\bool}{\code{bool}}
\newcommand{\stringg}{\code{string}}

\newcommand{\fresht}{\code{fresh\_t}}
\newcommand{\freshr}{\code{fresh\_r}}
\newcommand{\trans}{\code{trans}~}
\newcommand{\select}{\code{select}~}
\newcommand{\lett}{\code{let}~}
\renewcommand{\do}{\code{do}~}
\newcommand{\for}{\code{for}~}
\newcommand{\vphi}{\varphi}
\newcommand{\C}{\mathcal C}
\newcommand{\D}{\mathcal D}
\newcommand{\R}{\mathcal R}

\newcommand{\regle}{\cfrac}
\newcommand{\TDA}{ADT}

\section{Introduction}

    Pattern-matching on algebraic data-types (\TDA) allows the
    definition of functions by cases, a restricted form of rule based
    programming that is both relevant and powerful to specify function
    acting on \TDA s.  
 ML adopted a restricted form of pattern matching, where only the
    top-level structure of an \TDA\xspace is matched against the
    pattern~\cite{wadler}. Examples of more expressive patterns are
    given, e.g., by the Mathematica language.  
 However, both ML-like language or Mathematica are restricted to the
    handling of terms, that is, tree-shaped data structures (sets or
    multisets handled in Mathematica are represented by terms modulo
    associativity and commutativity).  

     In \cite{rule02} and~\cite{umc02} a framework where pattern
     matching can be expressed uniformly on many different data
     structures is exhibited. They rely on the notion of
     topological collection which embeds a neighborhood relation over
     its elements. The neighborhood relation enables the
     definition of a general notion of path (a sequential
     specification of a sub-structure); a pattern is used to specify a
     path that selects an arbitrary sub-collection to be substituted.
     This leads to a general functional language where the pattern
     matching is not limited to \TDA s.

We show in this paper that the topological collections bring a smooth
extension of the Hindley-Milner type
system~\cite{hindley}\cite{milner} with some polytypism
\cite{polytypism} and we suggest an extension of the Damas-Milner type
inference algorithm that allows to find a type to programs expressed
in an extension of mini-ML with topological collections and rule based
transformations over them.

Section~\ref{short} gives a brief description of the topological collections
and their transformation; section~\ref{types} gives an overview of
types in this framework; the types are investigated in
section~\ref{formel} where the typing rules and the inference
algorithm are given; several direct extensions of the language are
discussed in section~\ref{extensions} and section~\ref{conclusion}
concludes this paper.

\section{Topological Collections and Transformations}

Topological collections are data structures corresponding conceptually
to a mapping from a set of positions into a set of values such that
there is a neighborhood relation over the positions. Two values of a
collection are said to be neighbors if their positions are
neighbors. The \emph{sequence} is an example of topological collection
where the elements have at most a \emph{left} neighbor and a
\emph{right} neighbor. The NEWS \emph{grid} which is a generalization of
arrays of dimension 2 is another example where each element has at most
four neighbors, considering a Von Neumann neighborhood~\cite{lisper93}.

\label{short}

The notion of neighborhood is a means to embed in the programming
language the spatial locality of computations of programs.

Many other data structures can be seen from the topological point of
view. For example the \emph{set} and the \emph{multi-set} (or
\emph{bag}) are topological collections where each element is
neighbor of each other element (the set of positions of a set, is the
set of the elements itself). See~\cite{Aut01} for other examples of
topological collections.

These data structures come with a rule based style of programming: a
rule defines a local transformation by specifying some elements to be
matched and the corresponding action. The topological disposition
of the matched elements is expressed directly within the pattern of the
rule. Thus a collection can be transformed by the simultaneous
application of local transformations to non-intersecting matching
sub-sets of the collection.\\

The \MGS programming language described in~\cite{Aut01}
and~\cite{umc02} supplies the topological collections as first-class
values and transformations as a means to describe rule based functions
over collections. The language we work on in our paper is largely
inspired by \MGS although some features such as the possibility for a
collection to contain elements of different types have been left out.

In the rest of this section we describe the handling of collections
via rules in our restriction of \MGS.\\

A \emph{rule} is written $p \code{=>} e$ where $p$ is the pattern and
$e$ is the expression that will replace the instances of $p$. A
\emph{transformation} is a list of rules introduced by the keyword
\code{trans}.
The application of a transformation
$\code{trans[}p_1\code{=>}e_1;~p_2\code{=>}e_2\code]$ to a collection
$c$ consists in selecting a number of non-intersecting occurrences of
$p_1$ in $c$ such that there is no further possible occurrence; then
replacing the selected parts by the appropriate elements calculated from
$e_1$; then selecting a number of non-intersecting occurrences of $p_2$
and replacing them with the appropriate values.\\

The pattern can be a single element $x$ or a single element satisfying
a condition $x/e$ where $e$ is a boolean expression; it can also be a
two elements pattern $x,y$ such that $y$ is a neighbor of $x$. Here the comma
expresses the neighborhood relation and is not intended to express
a tuple. The pattern $x/(x=0), y/(y=1), z/(z=2)$ matches three values
such that the first is a 0, the second is a 1, the third is a 2, the
second is in the neighborhood of the first and the third is in the
neighborhood of the second.

The right hand side of the rule is composed of an expression denoting
the elements replacing the selected elements. In order to allow the
replacement of parts by parts of different size, the value expressed
in the right hand side of a rule must be a sequence. The elements of
this sequence will substitute the matched elements.  Thus we can
consider rules replacing sub-parts constituted of a single element
with several element, or sub-parts constituted of several elements
with one element or even with no element, and so on.

A way of building a sequence is using the empty sequence
\code{empty\_seq} and the constructor \code{::}. The syntactic
shortcut \code[$e$\code] can be used to express
$e$\code{::empty\_seq}.

\subsection{Two examples}

The following two examples show two programs acting
respectively on sequences and sets.

\paragraph{Sorting a Sequence.}

A kind of bubble-sort is immediate:\\

\code{trans[ x, y/(y<x) => y :: x :: empty\_seq ; x => [x] ]}\\

This two rules transformation has to be applied on the sequence until
a fixpoint is reached. The fixpoint is a sorted sequence.

This is not really the bubble-sort because the swapping of elements can happen
at arbitrary places; hence an out-of-order element does not
necessarily bubble to the top in the characteristic way.

We will see in section~\ref{formel} that the rule \code{x => [x]} is
required.

\paragraph{Eratosthene's Sieve on a Set.}

The idea is to apply the transformation on the set of the integers
between 2 and $n$. The transformation replaces an $x$ and an $y$ such
that $x$ divides $y$ by $x$. The iteration until a fixpoint of this
transformation results in the set of the prime integers less than
$n$.\\

\code{trans [ x, y/(y mod x = 0)  => [x] ; x => [x] ] }

\section{Typing the Collections and the Transformations}
\label{types}

The type of a topological collection is described by two pieces of
information: the type of the elements inside the collection and its
organization. The former is called its \emph{content type} and the
latter its \emph{topology} (see~\cite{jay95} for an example of
separation between the shape and the data). For example, a set of
integers and a set of strings do not have the same content type but
have the same topology. Collection types will be denoted by
$\col{\rho}{\tau}$ where $\tau$ is the content type and $\rho$ is the
topology.  Thus a set of strings will have the type
$\col{\set}{\stringg}$.

The usual notion of polymorphism of ML languages is provided on the
content type. For example the \emph{cardinal} function that returns the
number of elements of a set would have the type $\col{\set}{\alpha}
\fleche \int$ where $\alpha$ is a free type variable since it can be
applied to a set irrespectively of the type of its elements. The
nature of the content type does not affect the behavior of the
cardinal function, therefore the polymorphism is said to be uniform on
the content type.

Instead of providing  different functions that count the number of
elements for each topology, the language provides the function
\emph{size} with the type $\col{\theta}{\alpha} \fleche \int$ where
$\theta$ is a free topology variable. Functions that accept any kind
of topology are said to be \emph{polytypic}~\cite{polytypism}.\\

A way of handling collections is using polytypic operators and
constant collections: the constructor operator \code{::} has the type
$\alpha \fleche \col{\theta}{\alpha} \fleche \col{\theta}{\alpha}$;
the destructors \code{oneof} and \code{rest} have the type
$\col{\theta}{\alpha} \fleche \alpha$ and $\col{\theta}\alpha \fleche
\col\theta\alpha$ and are such that for any collection $c$,
\code{oneof($c$)} and \cod{rest($c$)} make a partition of
$c$ (see~\cite{buneman95}).

The constant collections are \cod{empty\_set}, \cod{empty\_seq} and so
on.\\

Collections can also be handled with transformations. As seen in the
previous section, transformations are functions on collections
described by rewriting rules. This kind of function is introduced by
the keyword \cod{trans}. For example the function \cod{trans [ x=>[x]
]} implements the identity over collections and has the type
$\col{\theta}\alpha \fleche
\col\theta\alpha$. It is the identity because it maps the identity to
all the elements of the collection.

As we said, the right hand side of a rule must be a sequence because
the pattern matched can be replaced by a different number of
elements. On some topologies such as the \emph{grid}, the pattern and
the replacement sequence must have the same size. If the sizes are not
compatible a \emph{structural error} will be raised at execution
time. These structural errors are not captured by our type system.
See~\cite{Aut01} for more details on the substitution process in the
collections.\\

The \emph{map} function can be expressed as follows:\begin{center}\cod{fun f
-> trans [ x => [f x] ]} \end{center}and has the type $(\alpha \fleche \beta) \fleche
\col\theta\alpha \fleche \col\theta\beta$.\\

Unlike in the original \MGS language, a collection cannot contain
elements of different types. We have chosen to set this restriction to allow
to build an inference algorithm in the Damas-Milner
style~\cite{damasmilner}. Allowing such heterogeneous collections
would lead to a system with subsumption and union types that would
need complex techniques to determine the types of a program.

\section{The Language}
\label{formel}

In this section we first describe the syntax of the studied
language. Then we describe the type verification rules and finally we
give the type inference algorithm that computes the principal type of
a program.

\subsection{Syntax}

Topological collections are values manipulated with constants,
operators, functions and transformations, no new syntactic
construction is needed.

For the transformation we have to enrich the syntax of
mini-ML~\cite{miniML} as shown in figure~\ref{syntaxe}.\\

\begin{figure}
\begin{center}

\begin{boxedminipage}{11cm}
\begin{tabular}{lrl}
$e$ & ::= & $id~|~cte ~|~(e,e)~|~ \code{fun }x\code{-> }e $\\
    & $|$ & $ e~e~|~\code{let }id=e\code{ in } e$\\
    & $|$ & $\code{trans} \code{ [} ~l~  \code]$\\
    &     &  \\ 
$l$ & ::= & $id=>e$ \\
    & $|$ & $p=>e \code{ ; } l$\\

\end{tabular}
\hfill
\begin{tabular}{lrl}
$ p $ 	& ::= & $id$ \\
	& $|$ & $id \code/ e$\\
      	& $|$ & $id \code{,} p$\\
	& $|$ & $id\code/e\code,p$\\
\end{tabular}
\end{boxedminipage}
\caption{Syntax of the language}
\label{syntaxe}
\end{center}
\end{figure}

The construction $p=>e$ is called a \emph{rule} and a transformation
is a syntactic list of rules. In the construction $id/e$ occurring in a
pattern, $e$ is called a \emph{guard}.

The last rule of a transformation must be a variable for exhaustiveness
purpose. Putting the rule \code{x => [x]} in last position of a
transformation expresses that all unmatched values are left
unmodified.
It is not possible to infer a relevant default case for a
transformation. For example the rule \code{x => [x]} cannot be the
default case for a transformation of the type $\col\theta{string} \fleche
\col\theta{int}$. Therefore the default case must be specified
explicitly by the programmer.
This explains the grammar for the list of rules $l$ which enforces the
presence of a last rule of the form $id => e$ matching every remaining
element. The expression $e$ in the right hand side provides the
appropriate default value.

We will use some operators such as \cod{::} in an infix position but
this syntax can be easily transformed into the one of figure~\ref{syntaxe}.
Operators are functional constants of the language. 

\subsection{The Type System}
\label{typesystem}

\subsubsection*{Types Algebra}
We enrich the polymorphic type system of mini-ML with the topological
collections.  The collection type introduces a new kind of
construction in types: the \emph{topology}.

From a type point of view, transformations are just functions that
act on topological collections without changing their topology, so no
new construct is needed for them in the type algebra.\\

\begin{tabular}{lrll}
Types~: & $\tau$  ::= &
 $T$ & base type (\code{int}, \code{float}, \code{bool}, \code{string})\\
       & $|$ & $\alpha$ & type variables\\
       & $|$ & $\tau \fleche \tau$ & functions\\
       & $|$ & $\tau \times \tau$  & tuples\\
       & $|$ & $\col{\rho}{\tau}$  & collections\\
       &   &                     & \\
Topologies~ : & $\rho$ ::= & 
 $R$ & base topology (\code{bag}, \code{set}, \code{seq}, \code{grid}, ...)\\
       & $|$ & $\theta$ & topology variables\\
\end{tabular}\\

We give in appendix~\ref{annexe} the definitions of $\FVt$ and $\FVr$ which
calculate the type variables and the topology variables occurring in a
type.

\subsubsection*{Type Schemes}

A type scheme is a type quantified over some type variables and some
topology variables:
$$ \sigma ::=
\forall[\alpha_1,\dots,\alpha_n][\theta_1,\dots,\theta_m].\tau$$

A type $\tau$ is an instance of a type scheme $\sigma=\forall
[\alpha_1,\dots,\alpha_n][\theta_1,\dots,\theta_m].\tau'$ and we write
$\sigma \leq \tau$ if and only if there are some types $\tau_1,\dotsc,\tau_n$
and some topologies $\rho_1,\dotsc,\rho_m$ such that 
$\tau=\tau'
[\alpha_1\shortleftarrow \tau_1,\dotsc,\alpha_n \shortleftarrow \tau_n,
 \theta_1\shortleftarrow \rho_1,\dotsc,\theta_m \shortleftarrow \rho_m]$
.\\

In the following, an environment is a function from identifiers
to type schemes.

$TC$ is the function that gives the type scheme of the constants of
the language. For example $TC(\cod{::})$ is $\forall[\alpha][\theta].\alpha \fleche \col\theta\alpha \fleche \col\theta\alpha$.

$\FVt$ and $\FVr$ are extended to type schemes and calculate the free
variables of a type scheme, that is the variables occurring in the type scheme which are not bound by the quantifier. For example if $\sigma$ is $\forall[\alpha_1][\theta_1].\col{\theta_1}{\alpha_1}\fleche\col{\theta_2}{\alpha_2} $ then $\FVt(\sigma)$ is $\alpha_2$ and $\FVr(\sigma)$ is $\theta_2$.

\subsubsection*{Typing Rules}

The typing rules are nearly the same as the Hindley-Milner rules~\cite{hindley}\cite{milner}. The
differences are that a rule has been added for the transformations and
that the notions of instance and the $Gen$ function have been adapted
to the type algebra.

The \emph{Gen} function transforms a type into a type scheme by
quantifying the variables that are free in the type and that are not
bound in the current environment. The definition of \emph{Gen} is
the following:

$Gen(\tau,\Gamma)~=~\forall
[\alpha_1,\dotsc,\alpha_n][\theta_1,\dotsc,\theta_m].\tau $ with $
\{\alpha_1,\dotsc,\alpha_n\}=\FVt(\tau)\backslash\FVt(\Gamma) $ and $
\{\theta_1,\dotsc,\theta_m\} =\FVr(\tau)\backslash\FVr(\Gamma)$.\\

The typing rules are:

$$\regle{
\Gamma(x) \leq \tau 
}{
\Gamma \infere x : \tau}~(var-inst)
\quad\quad
\regle{
TC(c) \leq \tau
}{
\Gamma \infere c:\tau}~(const-inst)
\quad\quad
%
$$

$$\regle{
\Gamma\union\{x:\tau_1\} \infere e : \tau_2
}{
\Gamma \infere (\cod{fun } x \fleche e) : \tau_1 \fleche \tau_2}~(fun)
\quad\quad
\regle{
\Gamma \infere e_1 : \tau' \fleche \tau \quad \Gamma \infere e_2 : \tau'
}{
\Gamma \infere e_1~e_2 : \tau}~(app)$$

$$\regle{
\Gamma \infere e_1 : \tau_1 \quad
\Gamma\union\{x:Gen(\tau_1,\Gamma)\} \infere e_2:\tau_2
}{
\Gamma \infere (\cod{let } x = e_1 \cod{ in } e_2) : \tau_2}~(let)$$

\small
$$\regle{
\dessusdessous{
\big\{ \Gamma \union \{ x_i^j:\tau\}_{(j\leq m_i)} \union\{\cod{self}:\col\rho\tau\} \infere e_i : \col{seq}{\tau'} \big\}_{(i\leq n)}
}{
\big\{ \Gamma \union \{x_i^j : \tau\}_{(j \leq k)} \union\{\cod{self}:\col\rho\tau\} \infere e_i^k : \bool \big\}_{(i\leq n),(k \leq m_i)}
}
}{
\Gamma \infere \trans \code[ 
x_1^1/e_1^1 \code, ... \code, x_1^{m_1}/e_1^{m_1} \code{=>}
	 e_1 \code; ... \code;
x_n^1/e_n^1 \code, ... \code, x_n^{m_n}/e_n^{m_n} \code{=>} e_n \code] : \col{\rho}{\tau} \fleche \col{\rho}{\tau'}
}(trans)$$
\normalsize

In the (trans) rule, $k_n$ is always equal to 1 and $e_n^1$ is always
equal to \code{true}.

Inside a rule the \cod{self} identifier refers to the collection the
transformation is applied on.

The (trans) rule expresses that a transformation has the type
$\col{\rho}{\tau} \fleche \col{\rho}{\tau'}$ if when you suppose that
all the $x_i^j$ have the same type $\tau$ and that $self$ has the type
$\col{\rho}{\tau}$ it can be proven that the $e_i^j$ are boolean values and
that the $e_i$ have the type $\col{seq}{\tau'}$.

We can see that if \cod{self} is not used in a transformation, this
one will be polytypic since $\rho$ will not be bound to any topology.

The following examples show a type verification on a polytypic
transformation and on a non-polytypic one.

\subsubsection*{Polytypic Example}

The following transformation can be proven to be an $\col\theta{int} \fleche \col{\theta}{int}$ function for any topology $\theta$.\begin{center}
\code{trans [ x, y/x>y => x :: y :: (x-y) :: empty\_seq ; x => [x] ]}
\end{center}

\newcommand{\racine}{\infere 
	\code{trans [ x,y/x>y => x::y::(x-y)::empty\_seq ; x=>[x] ]} :
	 \col\theta{int} \fleche \col{\theta}{int}}

\newcommand{\rl}{\Gamma_0 \infere \code{x>y} : bool}

\newcommand{\rr}{\Gamma_0 \infere \code{ x::y::(x-y)::empty\_seq} : \col{seq}{int}}

\newcommand{\rrl}{\Gamma_0 \infere \cod x :int}
\newcommand{\rrr}{\Gamma_0 \infere  \code{ y::(x-y)::empty\_seq} : \col{seq}{int}}

\newcommand{\rR}{\Gamma_1 \infere \cod{[x]} : \col{seq}{int}}
\newcommand{\rRc}{\Gamma_1 \infere \cod x : int}
\newcommand{\rRcc}{\Gamma_1(\cod x) \leq int}

\begin{figure}[htbp]
\begin{center}

\begin{sideways}
\subfigure{
$
\regle{
\regle{\dots}{\rl} 
\quad
\regle{\regle{\Gamma_0(\cod x) \leq int}{\rrl} \quad \regle{\dots}{\rrr}}{\rr}
\quad
\regle{\regle{\rRcc}{\rRc}}{\rR}
}
{
\racine
}
$
}
\end{sideways}
\begin{sideways}
\subfigure[]{
$
\regle{
	\regle{ \regle{TC(\cod{not})\leq bool \fleche bool}{\Gamma_2 \infere \code{not} : bool \fleche bool} 
	\quad
	\regle{\dots}{\Gamma_2 \infere \code{not(is\_left x self)} : bool} } {
	\Gamma_2 \infere \code{not(is\_left x self)} : bool } \quad
	\regle{ \regle{\dots}{\Gamma_2 \infere \code{x+(left x self)} :int} } {
	\Gamma_2 \infere \code{[x+(left x self)]} : \col{seq}{int} } 
\quad
\regle{\regle{\Gamma_2 (\cod x) \leq int}{\Gamma_2 \infere \cod{x} :int}}{\Gamma_2 \infere \cod{[x]} : \col{seq}{int}}
}
	{
\infere \code{trans [ x/(not(is\_left x self))=>[x+(left x self)] ; x=>[x] ]} : \col{seq}{int} \fleche \col{seq}{int}
}$
}

\end{sideways}

\caption{Two examples of type verification}
\label{tutu}

\end{center}	

\end{figure}

The proof is given in figure~\ref{tutu}a where
$\Gamma_0=\{x:int;y:int;self:\col\theta{int}\}$,
$\Gamma_1=\{x:int;self:\col\theta{int}\}$
 and with the following lemmas:
$$\regle{\Gamma \infere e_1 : int \quad \Gamma\infere e_2 :
\col{seq}{int}}{\Gamma\infere e_1 \code{::} e_2 : \col{seq}{int}}
\quad
\regle{\Gamma \infere e : \tau}{\Gamma \infere \code[ e \code] : \col{seq}\tau}
$$

\subsubsection*{Non-Polytypic Example}

The operator \cod{is\_left} acts as a predicate that returns
\emph{true} if the element is at the left extremity of the
sequence. Thus it returns \emph{false} is the element has a left
    neighbor. It can be used only within a transformation\footnote{The
    \cod{is\_left} operator is only available in transformations, where
    the identifiers introduced by the pattern are bound to a position in the
    collection. Allowing only such identifiers to be arguments of
    \code{is\_left} allows to remove any ambiguity on the position
    denoted in the sequence, even if the position contains a value
    occurring several times.} and takes two arguments: the first is a
    pattern variable and the second is a collection.
Similarly, the operator \cod{left} takes a pattern variable
    $x$ and a sequence $s$ and returns the left neighbor of $x$ in
    $s$.\\

Let us consider the following transformation:\begin{center}
\code{trans [ x/(not (is\_left x self)) => [x+(left x self)] ; x=>[x] ]}
\end{center}
This transformation does not have the same effect as the following one:
\begin{center}
\code{trans [ l, x => (l :: l+x :: empty\_seq) ; x=>[x] ]}
\end{center}
because in the former, every element $x$
of the sequence except the leftmost one will be replaced by the sum
of itself and its left neighbor whereas in the latter, the $l$ element
will be replaced by itself and thus will not be increased. For example
the former transformation applied to the sequence
\code{(1::2::3::4::empty\_seq)} results in
\cod{(1::3::5::7::empty\_seq)} whereas the application of the latter
transformation to the same sequence would result in
\code{(1::4::3::7::emty\_seq)}.

The figure~\ref{tutu}b where $\Gamma_2 = \{x:int;self:\col{seq}{int}
\}$ proves that the first transformation has the type $\col{seq}{int}
\fleche \col{seq}{int}$.

This transformation cannot be proven to have the type
$\col{\rho}{int}\fleche\col{\rho}{int}$ if $\rho \not = seq$ because
\cod{left} and \cod{is\_left} act exclusively on sequences.

\subsection{Type Inference}

The typing rules given in section~\ref{typesystem} are a means to
verify that a program has a given type but this type is a parameter of
the verification procedure. We now give the equivalent of the
Damas-Milner type inference that enables the full automated type
verification since it computes the principal type of a program. The
resulting type is said to be principal because every type that can
fit the program is an instance of this type.

The type inference algorithm is given after the unification procedure.

\subsubsection*{Unification}

\newcommand{\mgu}{\code{mgu}}

Unifying two types $\tau_1$ and $\tau_2$ consists in finding a
substitution $\varphi$ over the free variables of $\tau_1$ and
$\tau_2$ called the unifier such that
$\varphi(\tau_1)=\varphi(\tau_2)$.

A substitution is a most general unifier (mgu) for two types $\tau_1$
and $\tau_2$ if for any unifier $\varphi_1$ of $\tau_1$ and $\tau_2$,
there is a substitution $\varphi_2$ such that $\varphi=\varphi_2 \rond
\varphi_1$.

We give the $\mgu$ function that computes the most general unifier
of a set of pairs of types denoted by $\tau_1=\tau_2$. This function is
necessary to the type inference procedure. If \mgu\xspace fails then
there is no unifier for the given types.

The difference between our $\mgu$ and Damas and Milner's original
\emph{mgu} is the addition of the case for the collection types.
Two collection types are unified by unifying their content
types and their topologies. The substitution doing this unification is
found as $\varphi_1 \rond \varphi_2$ where $\varphi_2$ unifies the topologies
and $\varphi_1$ unifies the content types. The computation of $\varphi_2$ is
made by the dedicated $\cod{mgu}_r$ function. This function fails when
the two topologies are different base topologies since they cannot be
unified. The substitution $\varphi_2$ is applied to the content types
before computing $\varphi_1$ with \cod{mgu}.

The standard cases of the definition of \cod{mgu} are:\\
\noindent\begin{tabular}{rcl}
$\mgu\big( \emptyset\big)$ & = & $[\;]$\\
$\mgu\big( \{\tau = \tau\} \union C)$ & = & $\mgu(C)$\\
$\mgu\big( \{ \alpha = \tau \} \union C)$ (if $\alpha$ is not free in $\tau$)& = &
	let $\varphi = [\alpha \shortleftarrow \tau]$ in
	 $\mgu(\varphi(C)) \rond \varphi$\\
$\mgu\big( \{ \tau = \alpha \} \union C)$ (if $\alpha$ is not free in $\tau$)& = &
	let $\varphi = [\alpha \shortleftarrow \tau]$ in
	 $\mgu(\varphi(C)) \rond \varphi$\\
$\mgu \big(\{\tau_1 \fleche \tau_2 = \tau'_1 \fleche \tau'_2 \} \union C \big)$
	& = & $\mgu\big(\{\tau_1 = \tau'1 ~;~ \tau_2 = \tau'2 \} \union C\big)$\\
$\mgu \big(\{\tau_1 \times \tau_2 = \tau'_1 \times \tau'_2 \} \union C \big)$
	& = & 
	$\mgu\big(\{\tau_1 = \tau'1 ~;~ \tau_2 = \tau'2 \} \union C\big)$\\
\end{tabular}\\

\noindent The new case for the collections is:\begin{center}
\mbox{$\mgu\big( \{ \col{\rho}{\tau} = \col{\rho'}{\tau'}\} \union C \big)$  ~~=~~
	let $\varphi = \mgu_r (\rho = \rho')$ in 
	$\mgu\big( \varphi \big(\{ \tau = \tau' \} \union C \big)\big) \rond \varphi$}\end{center}

\noindent The unification of topologies is defined by:
\begin{center}
\begin{tabular}{rcl}
$\mgu_r( \rho = \rho )$ & = & $[\; ]$\\
$\mgu_r( \theta = \rho   )$ & = & $[\theta \shortleftarrow \rho]$ \\
$\mgu_r( \rho = \theta   )$ & = & $[\theta \shortleftarrow \rho]$ \\
\end{tabular} 
\end{center}

\subsubsection*{Type Inference}
The type reconstruction algorithm is nearly the same as the Damas-Milner
one. The differences are that it uses specialized versions of
\emph{mgu} and \emph{Gen} functions and that there is a new case for the
transformations. It is described here in an imperative way: $\varphi$
is the current substitution and $V_t$ and $V_r$ are sets of free
type variables and topology variables.

The algorithm is given in figure~\ref{inference}.

\begin{figure}

\begin{center}
\begin{boxedminipage}{14cm}
\setlength{\leftmargin}{1cm}
\small
\begin{center}
\begin{tabular}{rll}
\fresht	& = & let $\alpha \in V_t$\\
		&   & do $V_t \shortleftarrow V_t \backslash \{ \alpha \}$\\
	  	&   & return $\alpha$\\

\freshr	& = & let $\theta \in V_r$\\
		&   & do $V_r \shortleftarrow V_r \backslash \{ \theta \}$\\
	  	&   & return $\theta$\\
\end{tabular}\\
\end{center}

$W(\Gamma \infere e) =$\\
\cod{(* original cases *)}\\
If $e= x$\\
$~\quad$let $\forall [\alpha_1, \dots , \alpha_n][\theta_1,...,\theta_m].\tau = \Gamma(x)$\\
$~\quad$let $\alpha'_1, \dots,\alpha'_n=\fresht,\dots,\fresht$\\
$~\quad$let $\theta'_1, \dots,\theta'_m=\freshr,\dots,\freshr$\\
$~\quad$return $\tau[\alpha_1 \shortleftarrow \alpha'_1, \dots ,
	 \alpha_n \shortleftarrow \alpha'_n,
	 \theta_1 \shortleftarrow \theta'_1, \dots,
	\theta_m \shortleftarrow \theta'_m]$\\	
If $e= \cod{fun } x \fleche e  $\\
$~\quad$let $\alpha$ = \fresht \\
$~\quad$let $\tau=W(\Gamma\union{x:\forall[\;][\;].\alpha} \infere e)$\\
$~\quad$return $\alpha \fleche \tau$\\
If $e= e_1~e_2$\\
$~\quad$let $\tau_1=W(\Gamma \infere e_1)$\\
$~\quad$let $\tau_2=W(\Gamma \infere e_2)$\\
$~\quad$let $\alpha$ = \fresht\\
$~\quad$do $\varphi \shortleftarrow \cod{mgu}( \varphi(\tau_1) =\varphi(\tau_2 \fleche \alpha)) \rond \varphi$\\
If $e=\cod{let } x = e_1 \cod{ in } e_2$\\
$~\quad$let $\tau_1 = W(\Gamma \infere e_1)$\\
$~\quad$let $\sigma = Gen(\varphi(\tau_1),\varphi(\Gamma))$\\
$~\quad$return $W(\Gamma\union\{x:\sigma\} \infere e_2)$
\begin{center}
\begin{boxedminipage}{13.5cm}
\cod{(*   new case for the transformations   *)}\\
If $e=\trans \cod[p_1 \code{=>} e_1 \code;~ ... \code;~
	p_n \code{=>} e_n\cod]$\\ 
   $~\quad$let $\alpha,\beta = \fresht , \fresht$\\
   $~\quad$let $\theta = \freshr$\\
   $~\quad\for i=1..n$\\
   $~\quad\quad$let $id_i^1/e_i^1,\dots,id_i^{m_i}/e_i^{m_i} = p_i$\\
$~\quad\quad\for j=1..m_i$\\
$~\quad\quad\quad $let $\tau_i^j = W\big(\Gamma \union\{ \self:\col{\theta}{\alpha}\}\union\{id_i^k : \alpha\}_{k\leq j} \infere e_i^j)$\\
$~\quad\quad\quad $do $\varphi \shortleftarrow \mgu \big( \{ \varphi(\tau_i^j) = bool\})
	\rond \varphi$\\
   $~\quad\quad$let $\tau_i = 
	W\big(\Gamma \union \{\self:\col{\theta}{\alpha}\}\union
	\{ id_i^k : \alpha \}_{k\leq m_i}
\infere e_i\big)$\\
   $~\quad\quad$do $\vphi \shortleftarrow
 \mgu\big( \{\vphi(\tau_i) = \vphi(\col{\seq}{\beta}) \}\big)
 \rond \vphi$\\
   $~\quad$return $\col{\theta}{\alpha} \fleche \col{\theta}{\beta}$
\end{boxedminipage}
\end{center}
\normalsize
\smallskip

\end{boxedminipage}
\caption{Type inference algorithm}
\label{inference}
\end{center}	
\end{figure}

The case for the transformations consists in unifying the types of all
the pattern variables and unifying the types of the right hand side
rules together and with a sequence collection type. These unifications
have to be made with respect to the guards that are boolean
values. 

If $W$ succeeds it computes the most general type of the program
analyzed and this one can be run without type error. If it fails
because of an \cod{mgu} or an $\cod{mgu}_r$ failure then the program is
ill-typed and might lead to a type error at execution time.

\section{Extensions}
\label{extensions}

\subsection{Repetition in a Pattern}

The \emph{star} \cod* expressing an arbitrary repetition of a
sub-pattern during the matching process has been introduced
in~\cite{rule02}. The pattern \code{ x/(x=0), * as y, z/(z=0)} for example
can match an arbitrary subcollection such that it contains two 0 and
that there is a \emph{path} between these 0. This means that one can
reach the second 0 from the first one only by going from an element to
one of its neighbors repetitively.

To take the star into account we modify the syntax of the patterns as
follows:\\
\begin{center}
\begin{tabular}{l}
 $p ~::=~ q ~|~ q,p$\\
$q ~ ::= ~ id ~|~ *~as~id$ 
\end{tabular}
\end{center}
where $q$ stands for elementary patterns.

We have not kept the guards in the elementary patterns in
order to keep the formulas readable but their addition does not lead to
new problems.

The elements matched by the star are named and can be referred to as a
sequence.

The star could have been considered as a repetition of a subpattern
as in \code{(x,y/x=y)*} but we have chosen to restrict the star to the
repetition of single elements for the sake of simplicity.\\

Before giving the new typing rule, we introduce a function which gives
the type binding corresponding to an elementary pattern: $b(q,\tau)$
is such that $b(x,\tau) = (x:\tau)$ and $b(\cod{* as }x,\tau) =
(x:\col{seq}{\tau})$. This function is used in the \emph{trans} typing
rule which is modified as follows:

$$\regle{
\big\{ \Gamma 
 \union \{ b(q_i^j,\tau) \}_{j \leq m_i} 
 \union\{\self :  \col{\rho}{\tau} \}
 \infere e_i : \col{\seq}{\tau'} \big\}_{i\leq n}
}{
\Gamma \infere \trans \code[ 
q_1^1 \code, ... \code, q_1^{m_1} \code{=>} e_1 \code; ... \code;
q_n^1 \code, ... \code, q_n^{m_n} \code{=>} e_n \code] : \col{\rho}{\tau} \fleche \col{\rho}{\tau'}
}~(trans')$$

\subsection{Directions in Patterns}

In section~\ref{typesystem} we saw the operator \code{left} that
returns the left neighbor of an element in a sequence. In the
framework of topological collections, a topology can supply several
neighborhood operators. For example \code{left} and \code{right} are
the neighborhood operators of the sequence and \code{north} and
\code{east} are neighborhood operators of the grid. Neighborhood operators are also called \emph{directions}.

A direction can be used to refine the patterns: the commas of the
pattern can be substituted by a direction to restrict the accepted
neighbors for the rest of the pattern. The substituting direction is
surrounded with the symbols \code| and \code> to sketch a kind of arrow.

For example if $d$ is a direction we can use the pattern \cod{x |d> y}
which is a shortcut\footnote{The expression \cod{y=(d x self)} in a guard where \cod{y} is a pattern variable and \cod d is a direction tests that the values denoted are the same and that their positions in the collection are the same. See the \MGS manual~\cite{mgslanguage} for more details.} for \cod{x,y/y=(d x self)}. However, the pattern
\cod{x |d> y} allows faster research of the instances of the pattern
in the collection than \cod{x,y/y=(d x self)}.

The pattern \cod{x |d> y} can be typed as \cod{x,y/y=(d x self)}.

\subsubsection*{The Bead-Sort Example}

\begin{figure}

\begin{center}
\subfigure[]{\includegraphics[scale=0.6]{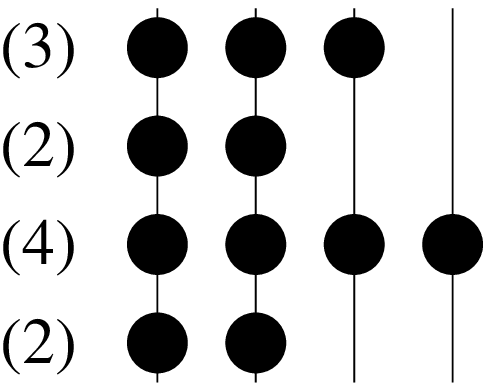}}\hspace{0.8cm}
\subfigure[]{\includegraphics[scale=0.6]{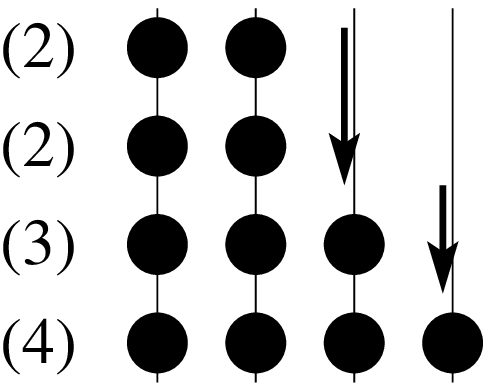}}\hspace{0.8cm}
\subfigure[]{\includegraphics[scale=0.6]{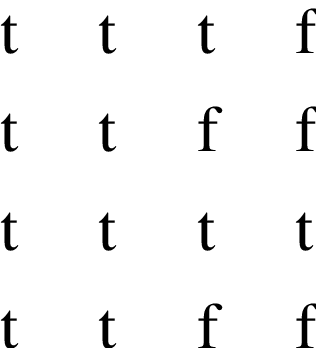}}\hspace{0.8cm}
\subfigure[]{\includegraphics[scale=0.6]{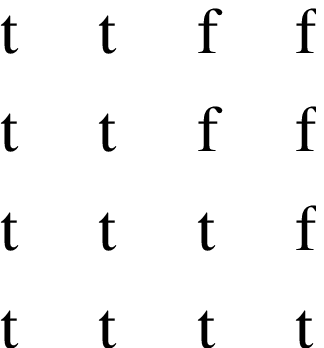}}
\end{center}
 
\caption{The Bead-Sort}
\label{boulier}
\end{figure}

The bead-sort is an original way of sorting positive integers presented
by~\cite{beadsort}. The sorting algorithm considers a column of numbers
written in unary basis. Figure~\ref{boulier}a shows the numbers 3, 2,
4 and 2 where the beads stand for the digits. The sorting is done by
letting the beads fall down as shown on figure~\ref{boulier}b.

The problem can be represented on a grid of booleans where \emph{true} stands
for a digit and \emph{false} for the absence of digit as shown on
figure~\ref{boulier}c. The bead-sort is achieved by iterating the application of the following transformation until a fixpoint is reached:\begin{center}
\cod{trans [ x/x=false |north> y/y=true => y::x::empty\_seq ; x=>[x] ]}
\end{center}
 
The first rule of this transformation is expressed as\begin{center}
\cod{x/x=false , y/(y=true \&\& y=north x self) => y::x::empty\_seq}
\end{center} in order to fit the type system. The result of $W$ on this transformation is $\col{grid}{bool} \fleche \col{grid}{bool}$.

\subsection{Strategies}
As far as the rules application strategy guarantees that every element
of the collection is matched (this is always possible since the last
rule always matches) the type system is not affected.

For instance, the \MGS language provides several strategies such as higher
priority given to the first rules or random application of the rules.

\section{Conclusion}
\label{conclusion}

Including the topological collections and pattern matching programming
on these structures in the ML framework allows to bring together a
powerful programming language with a rule programming framework common
to several other languages.

Our algorithm has been tested on \MGS programs and has been included
in a prototype \MGS compiler in order to achieve type-oriented
optimizations on the produced code. We believe that the best pattern
matching algorithms would be wasted on a dynamically typed language and
thus a type inference algorithm is an important step in the
development of an efficient compiler for rule based transformations.

However some restrictions on the \MGS language had to be done in order
to keep the simplicity of the Damas-Milner algorithm. We are currently
working on a type inference system with union
types~\cite{aiken-wimmers} to account for heterogeneous collections
supplied by the \MGS language.

Finally, we said that an error could occur when a transformation tries
to replace a subpart by a part of different shape on topologies as the
grid which cannot get out of shape. Such errors are not type errors
but some of them could be detected statically with a specific type
based analysis. Some research such as~\cite{jay95} manage with this
kind of error but the concerned languages do not provide the
flexibility of the rule based transformations proposed here.

\appendix
\section{Free Variables} 
\label{annexe}

The free variables of a type are the variables occurring in that
type. $\FVt$ computes the free type variables whereas $\FVr$ computes
the free topology variables.

\begin{tabular}{rcl}
$\FVt(T)$      & = & $\emptyset$\\
$\FVt(\alpha)$ & = & $\{\alpha \}$\\
$\FVt(\tau_1 \fleche \tau_2)$ & = & $\FVt(\tau_1) \union \FVt(\tau_2)$\\ 
$\FVt(\tau_1 \times \tau_2)$  & = & $\FVt(\tau_1) \union \FVt(\tau_2)$\\ 
$\FVt(\col{\rho}{\tau})$      & = & $\FVt(\tau)$\\
\end{tabular} \hfill \begin{tabular}{rcl}
$\FVr(T)$      & = & $\emptyset$\\
$\FVr(\alpha)$ & = & $\emptyset$\\
$\FVr(\tau_1 \fleche \tau_2)$ & = & $\FVr(\tau_1) \union \FVr(\tau_2)$\\ 
$\FVr(\tau_1 \times \tau_2)$  & = & $\FVr(\tau_1) \union \FVr(\tau_2)$\\ 
$\FVr(\col{\theta}{\tau})$    & = & $\{\theta \}  \union \FVr(\tau)$ \\
$\FVr(\col{R}{\tau})$         & = & $\FVr(\tau)$ \\

\end{tabular}

The free variables of a type scheme are the non-quantified variables
occurring in it:
$$ \FVt(\forall [\alpha_1,\dots,\alpha_n],[\theta_1,\dots,\theta_m].\tau) =
 \FVt(\tau) \backslash \{\alpha_1,\dots,\alpha_n\}$$
$$ \FVr(\forall [\alpha_1,\dots,\alpha_n],[\theta_1,\dots,\theta_m].\tau) =
 \FVr(\tau) \backslash \{\theta_1,\dots,\theta_m\}$$

\bibliographystyle{entcs}

\bibliography{BIB/biblio}

\begin{thebibliography}{10}
\expandafter\ifx\csname url\endcsname\relax
  \def\url#1{\texttt{#1}}\fi
\expandafter\ifx\csname urlprefix\endcsname\relax\def\urlprefix{URL }\fi
\newcommand{\enquote}[1]{``#1''}

\bibitem{aiken-wimmers}
Aiken, A. and E.~Wimmers, \emph{Type inclusion constraints and type inference},
  in: \emph{Proceedings of the Seventh ACM Conference on Functional Programming
  and Computer Architecture}, 1993, pp. 31--41.

\bibitem{beadsort}
Arulanandham, J.~J., C.~S. Calude and M.~J. Dinneen, \emph{Bead-{S}ort: A
  natural sorting algorithm}, EATCS Bull \textbf{76} (2002), pp.~153--162.

\bibitem{buneman95}
Buneman, P., S.~Naqvi, V.~Tannen and L.~Wong, \emph{Principles of programming
  with complex objects and collection types}, Theoretical Computer Science
  \textbf{149} (1995), pp.~3--48.

\bibitem{miniML}
Clement, D., J.~Despeyroux, T.~Despeyroux and G.~Kahn, \emph{A simple
  applicative language: {M}ini-{ML}}, in: \emph{Proceedings of the ACM
  conference on LISP and Functional Programming}, 1986, pp. 13--27.

\bibitem{damasmilner}
Damas, L. and R.~Milner, \emph{Principal type-schemes for functionnal
  programs}, in: \emph{Proceedings of the 15'th Annual Symposium on Principles
  of Programming Languages} (1982), pp. 207--212.

\bibitem{mgslanguage}
Giavitto, J.-L. and O.~Michel, \emph{{MGS}: a programming language for the
  transformations of topological collections}, Technical Report lami-61-2001,
  LaMI Universit\'e d'\'Evry Val d'Essonne (2001).

\bibitem{Aut01}
Giavitto, J.-L. and O.~Michel, \emph{{MGS}: a rule-based programming language
  for complex objects and collections}, in: M.~van~den Brand and R.~Verma,
  editors, \emph{Electronic Notes in Theoretical Computer Science},  ~
  \textbf{59} (2001).

\bibitem{umc02}
Giavitto, J.-L. and O.~Michel, \emph{Data structure as topological spaces}, in:
  \emph{Proceedings of the 3nd International Conference on Unconventional
  Models of Computation {UMC02}},  ~  \textbf{2509}, Himeji, Japan, 2002, pp.
  137--150, {L}ecture Notes in Computer Science.

\bibitem{rule02}
Giavitto, J.-L. and O.~Michel, \emph{Pattern-matching and rewriting rules for
  group indexed data structures}, in: \emph{{RULE'02}} (2002), pp. 55--66.

\bibitem{hindley}
Hindley, J., \emph{The principal type scheme of an object in combinatory
  logic}, Transactions of the American Mathematical Society \textbf{146}
  (1969), pp.~29--60.

\bibitem{jay95}
Jay, C.~B., \emph{A semantics for shape}, Science of Computer Programming
  \textbf{25} (1995), pp.~251--283.

\bibitem{polytypism}
Jeuring, J. and P.~Jansson, \emph{Polytypic programming}, in: J.~Launchbury,
  E.~Meijer and T.~Sheard, editors, \emph{{A}dvanced {F}unctional
  {P}rogramming, {S}econd {I}nternational {S}chool} (1996), pp. 68--114, {LNCS}
  1129.

\bibitem{lisper93}
Lisper, B. and P.~Hammarlund, \emph{On the relation between functional and
  data-parallel programming languages}, in: \emph{Proc. of the 6th. Int. Conf.
  on Functional Languages and Computer Architectures}, ACM, 1993, pp. 210--222.

\bibitem{milner}
Milner, R., \emph{A theory of type polymorphism in programming}, Journal of
  Computer and System Sciences \textbf{17} (1978), pp.~348--375.

\bibitem{wadler}
Wadler, P., \enquote{Efficient compilation of pattern matching,} Prentice-Hall,
  1987 Ch. 6 of "The Implementation of Functionnal Programming Language", S. L.
  Peyton Jones.

\end{thebibliography}

\end{document}